\documentclass[amsmath,amssymb,aps,twocolumn,physrev]{revtex4-2}

\usepackage{dcolumn}
\usepackage{bm}
\usepackage{pgfplots}
\usepackage{tikz}
\usepackage[mode=buildnew]{standalone}
\usepackage{graphicx}
\usepackage{color, colortbl}
\usepackage{comment}
\usepackage{float}
\usepgfplotslibrary{external}
\begin{document}

\title{Global Signals of the First Molecules from the Dark Ages in the Presence of Primordial Magnetic Fields
}

\author{Yu. Kulinich$^{1}$}
\email[]{yuriy.kulinich@lnu.edu.ua}
\author{B. Novosyadlyj$^{1, 2}$}
\author{M. Tsizh$^{1, 3}$}
\author{N. Fortuna$^{1}$}

\affiliation{$^{1}$ Astronomical Observatory of Ivan Franko National University of Lviv, Kyryla i Methodia str., 8, Lviv, 79005, Ukraine;}

\affiliation{$^{2}$International Center of Future Science and College of Physics of Jilin University,
2699 Qianjin Str., 130012, Changchun, P.R.China;}

\affiliation{$^{3}$Dipartimento di Fisica e Astronomia, Universitá di Bologna, Via Gobetti 92/3, 40121, Bologna, Italy}

\date{\today}

\begin{abstract}

We investigate how primordial magnetic fields (PMFs) affect the formation kinetics of the first molecules, H$_2$, HD, and HeH$^+$, as well as the populations of rovibrational levels and the global signals in the rovibrational transitions of H$_2$ and HD. For this purpose, we numerically solve a system of differential equations that describes the gas energy balance, the kinetics of the formation of the first molecules, and the populations of their rovibrational levels, taking into account the PMFs energy dissipation through ambipolar diffusion and decaying turbulence. We show that PMFs can significantly speed up the formation and destruction of the first molecules, leading to an increase in the number density of H$_2$ and HD molecules and a decrease in the number density of HeH$^+$ ion-molecules compared to the case without PMFs. We demonstrate that more frequent collisions of the gas particles in such models alter the ortho-to-para ratio of hydrogen molecules, making it a potential probe of the thermal history of gas in the early Universe. In contrast to the standard cosmological model, where the global signal from the first molecules appears as an absorption feature in the cosmic microwave background spectrum, cosmological models with PMFs can produce an emission signal. Specifically, for non-helical PMFs with $n_B = -2.9$ and a strength of $\sim 1$~nG, the signal transforms into emission with an amplitude of about $\sim 0.5$~Jy/sr. This signal is comparable in magnitude to other known CMB spectral distortions and falls within the detection capabilities of several proposed missions, including Super-PIXIE, Multi-SIMBAD (4 units), and Voyage2050. We show that both the amplitude and the spectral range of the global signals from the first molecules are highly sensitive to the spectral index $n_B$, the strength $B_0$, and the helicity of the PMFs. Therefore, the global signals from the first molecules can serve as a potential probe of PMFs. 

\end{abstract}


\maketitle

\section{Introduction} \label{intro}

The results of resent probes of the intergalactic magnetic field (IGMF) with high-energy ($\sim$TeV) gamma rays emitted by distant blazars suggest that the strength of IGMF at cosmological scales ($>$ 1 Mpc) is greater than $\sim 10^{-16}$--$10^{-18}$~G \citep{Neronov2010, Tavecchio2010, Tavecchio2011, Huan2011, Taylor2011, Dolag2011, Vovk2012, Chen2015}. So far, there is no generally accepted mechanism for generating magnetic fields in voids that fill most of the volume of the Universe during the formation of large-scale structures. It is natural to assume that IGMF originates from the primordial magnetic fields (PMFs) which arose in the early Universe as a result of some magnetogenesis process (see overview \cite{Batista2021}).
The strength of PMFs at cosmological scales should increase with redshift adiabatically, ${B} \sim(1 + z)^2$, with increasing effect on the ionized gas. Therefore, when approaching the last scattering surface, the interaction between PMFs and ionized gas in the intergalactic medium (IGM) should be taken into account. PMFs can dissipate their energy into the IGM through ambipolar diffusion and decaying turbulence \citep{Sethi2005}. These processes can substantially alter the thermal and ionization history of the post-recombination Universe \citep{Sethi2005, Chluba:2015lpa}, which would significantly affect the formation of the first molecules in the Dark Ages \citep{10.1111/j.1365-2966.2008.13302.x}. Here, we investigate such an effect using an updated model of primordial chemistry and considering two extreme cases of helical and non-helical PMFs. In this paper, we demonstrate that the presence of PMFs also affects the populations of the rovibrational levels of the first molecules, H$_2$ and HD, thereby modifying their expected signal signatures. For this purpose, we numerically integrate the system of differential equations describing the thermal and chemical evolution of the Universe and the kinetics of the rovibrational levels populations of the first molecules.

All computations in the paper were performed for consistent values of the main parameters of the cosmological model following the final data release of the Planck collaboration \citep{Planck2020}: the Hubble constant $H_0=67.4$ km/s/Mpc, the mean density of baryonic matter in the units of critical one $\Omega_b=0.0493$, the density parameter of dark matter $\Omega_{dm}=0.2657$, the density parameter of dark energy $\Omega_{de}=0.685$, its equation of state parameter $w_{de}=-1.03$, and the current temperature of cosmic microwave radiation $T_0=2.7255$ K. We also set the primordial helium abundance $Y_{\rm He}=0.2446$ \citep{Peimbert2016} and the deuterium fraction $y_{Dp}=2.527\cdot10^{-5}$ \citep{Cooke2018}, which are in good agreement with the posterior means from \cite{Planck2020}.

This article is organized as follows. In the next section (\ref{PMF}), we describe the statistical properties of PMFs using power spectra. Following that, in Section \ref{thermal} we carefully reproduce the evaluation carried out by others before on how the magnetic fields transfer their energy into gas in the time range between the last scattering and the Cosmic Dawn through processes of ambipolar diffusion and decaying turbulence. The energy transfer alters the thermal history of the Cosmic Dark Ages, which, in turn, changes the kinetics of primordial chemical reactions. The numerical computation of this kinetics is described in Section \ref{pchemistry}. After this, in Section \ref{fmolecules}, we assess the signal from the first molecules, which manifest themselves as the distortion of the CMB radiation spectrum. In the final \ref{concl} section, we discuss the possibility of observing these distortions with future missions and summarize the paper.

\section{Primordial Magnetic Fields description} \label{PMF}

For simplicity, PMFs are considered as a statistically isotropic Gaussian distributed random field with the Fourier space two-point correlation given by \cite{Pogosian:2002, Kunze:2010ys}
\begin{eqnarray*}
 \langle \tilde{B}^*_i (\mathbf{k}) \tilde{B}_j (\mathbf{k}') \rangle
&=& \frac12{(2\pi)^3} \delta_\mathrm{D} (\mathbf{k}-\mathbf{k}')\times \label{twopcor}\\
&&\left(P_{ij} P_S(k)+i\epsilon_{ijm}\hat{k}^mP_H(k) \right),\nonumber
\end{eqnarray*}
where $\tilde{B}_j(\mathbf{k})\equiv\int d^3xa^{i\mathbf{kx}}B_j(\mathbf{x})$, $\epsilon_{ijm}$ is the antisymmetric tensor and $P_{ij}\equiv \delta_{ij}-\hat{k}_i \hat{k}_j$ is the plane projector that satisfies $P_{ij}P_{jk} = P_{ik}$ and $P_{ij}\hat{k}^j = 0$ with unit wavenumber components $\hat{k}_i = k_i/k$. The symmetric part of the power spectrum, $P_S(k)$, is related to the energy density of the magnetic field as $\rho_{\rm mf} =(2\pi)^{-3}\int  dk k^2 P_S(k)/2$, and the antisymmetric part of the power spectrum, $P_H(k)$, is related to the helicity density of the magnetic field ${h}_{\rm mf} = (2\pi)^{-3}\int dk k|P_H(k)|/2$, 
which plays an important role in the efficiency of magnetic dynamos \cite{Brandenburg2005, Brandenburg2009}. The spectra have to satisfy the so-called realizability condition $|P_H(k)| \le P_S(k)$, which is a consequence of the Schwartz inequality when applied to the average helicity \cite{Brandenburg2005, Brandenburg2009, Kunze:2012, Kahniashvili:2014dfa}. Due to this, $P_S(k)$ is also called maximal (spectral) helicity, and the actual spectral helicity density may be expressed as a fraction $f_H(k)\equiv |P_H(k)|/P_S (k)$ of it, with $0 \le f_H (k)\le 1$. 
Since there are no estimations on the IGMF spatial distribution, it is common to set both power spectra of the PMFs as the power laws in $k$ space as $P_S(k) = A_Sk^{n_S}$ and $P_H(k) = A_H k^{n_H}$ respectively, where  $n_S>-3$ and $n_H>-4$ are spectral indexes of the symmetric and antisymmetric field components, which values depend on the mechanism that generates the PMFs. The amplitudes $A_{S}, A_{H}$ can be represented with rms values of the magnetic field $B$ and absolute value of the helicity $\mathcal{H}$ smoothed over a sphere of comoving radius $\lambda$ with a Gaussian filter function \citep{Ballardini_2015}
\begin{eqnarray}
        B_{\lambda}^2  &\equiv&  \int \frac{dk^3}{(2\pi)^3} P_Se^{-\lambda^2k^2}
        =  {A_S}\frac{k_{\lambda}^{n_S+3}}{(2\pi)^{n_S+5}}\Gamma\left(\frac{n_S+3}{2}\right), \nonumber\\
        \mathcal{H}_{\lambda}^2  &\equiv &  \int \frac{\lambda kdk^3}{(2\pi)^3}|P_H|e^{-\lambda^2k^2} 
        =  {A_H}\frac{k_{\lambda}^{n_S+3}}{(2\pi)^{n_H+5}}\Gamma\left(\frac{n_H+4}{2}\right). \nonumber
\end{eqnarray}
As a result, the symmetric and antisymmetric parts of the PMFs power spectrum are parameterized as 
\begin{eqnarray}
P_S(k) ={B_{\lambda}^2}\frac{(2\pi)^{n_S+5}}{\Gamma\left(\frac{n_S+3}{2}\right)}\frac{k^{n_S}}{k_\mathrm{\lambda}^{n_S+3}}, \,
P_H(k) ={\mathcal{H}_{\lambda}^2}\frac{(2\pi)^{n_H+5}}{\Gamma\left(\frac{n_H+4}{2}\right)}\frac{k^{n_H}}{k_\mathrm{\lambda}^{n_H+3}}.\nonumber
\end{eqnarray}
The energy density of PMFs at the moment of recombination can be obtained through mean square values of the magnetic fields at the damping scale $\lambda_{\rm D}^{\rm rec}$: 
\begin{eqnarray}
\rho_{\rm mf}^{\rm rec} \equiv \frac{|\mathbf{B}^{\rm rec}_\mathrm{\lambda_{\rm D}^{\rm rec}}|^2}{8\pi} = \frac{1}{2(2\pi)^3}\int \limits_{0}^{\infty} dk k^2 P_S(k)e^{-(\lambda^{\rm rec}_{\rm D}k)^2} \nonumber\\
\simeq\frac{|\mathbf{B}_{\lambda}^{\rm rec}|^2}{8\pi}
\left(\frac{k_\mathrm{D}^{\rm rec}}{k_{\lambda}}\right)^{n_S+3}.
\end{eqnarray}

The damping scale of PMFs at the moment of recombination is defined in \cite{Subramania1998, Subramanian_2016} as $
\lambda_{\rm D}^{\rm rec}\equiv2\pi/k_{\rm D}^{\rm rec}=2\pi\sqrt{{3}/{5}}(V_{\rm A}/c)\lambda^\gamma_{\rm D}$ for the damping factor of the form $\sim \exp(-k^2/k_{\rm D}^2)$. In our case, using the damping factor $\sim \exp(-k^2\lambda_{\rm D}^2)$, the damping scale must be redefined as $
\lambda_{\rm D}^{\rm rec}\equiv2\pi/k_{\rm D}^{\rm rec}=\sqrt{{3}/{5}}({V_{\rm A}}/{c})\lambda^\gamma_{\rm D}$. According to \cite{Kunze2014}, the magnetohydrodynamic Alf{\' v}en velocity at this epoch, expressed in units of the speed of light, is
\begin{eqnarray}
    \frac{V_{\rm A}}{c} &=&3.24\cdot10^{-4}\left(\frac{B^{\rm rec}_{\lambda_{\rm D}} \cdot (1+z_{rec})^{-2}}{\rm 1\, nG}\right)h^{-1}\nonumber\\&=&3.24\cdot10^{-4}\left(\frac{B_0}{\rm 1\, nG}\right)\left(\frac{ k_{\rm D}^{\rm rec}}{k_\lambda}\right)^{\frac{n_{\rm B}+3}{2}}h^{-1},\nonumber
\end{eqnarray}
where $B_0 \equiv B^{\rm rec}_{\lambda}\cdot(1+z_{rec})^{-2}$ is the rms of PMFs strength smoothed over scale $\lambda$ and recalculated to present time, $z=0$. The comoving Silk damping scale can be approximated by the formula \citep{Hu_Sugiyama_1996}:
\begin{equation}
\lambda^\gamma_{\rm D}\approx 5.7\omega_m^{-\frac{1}{4}}\omega_b^{-\frac{1}{2}}\left(\frac{x_e^{rec}}{0.1}\right)^{-\frac{1}{2}}\left(\frac{1+z_{\rm rec}}{1100}\right)^{-\frac{5}{4}}\,{\rm Mpc}. 
\end{equation}
This formula quantifies the comoving scale at which baryon acoustic oscillations are suppressed by photon-baryon diffusion, with the parameters representing the present-day matter density ($\omega_m=\Omega_mh^2$), Hubble constant ($h=100$~km/s/Mpc), baryon density ($\omega_b=\Omega_bh^2$), ionization fraction ($x_{\rm e}^{rec}$), and the redshift at recombination ($z_{\rm rec}$). 
Finally, the damping scale of PMFs at the epoch of cosmological recombination can be approximated as \citep{Novosyadlyj:2025zxs}
\begin{eqnarray}
\hskip-0.5cm k_{\rm D}^{rec}&\approx &\left[7.3\cdot10^4\left(\frac{k_\lambda}{{\rm Mpc}^{-1}}\right)^{n_{\rm B}+3}\left(\frac{\omega_{\rm m}}{0.15}\right)^{\frac{1}{2}}\left(\frac{\omega_{\rm b}}{0.02}\right)\right. \\
\hskip0.5cm&\times&\left(\frac{x_{\rm e}^{rec}}{0.1}\right)\left.\left(\frac{1+z_{\rm rec}}{1100}\right)^{\frac{5}{2}}\left(\frac{\rm nG}{B_{0}}\right)^2\left(\frac{h}{0.7}\right)^2\right]^{\frac{1}{n_{\rm B}+5}} \, {\rm Mpc}^{-1},\nonumber
\label{kDrec}
\end{eqnarray}
The upper limits for PMFs strength vary depending on the cosmological probe used, but they generally fall in the nanogauss (nG) range. There are several constraints on the upper limit for the non-helical PMFs strength $B_0$ smoothed over scale $\lambda = 1$~Mpc. Planck data constrain the amplitude of PMFs at scale $\lambda = 1$~Mpc to less than a few nG \citep{Plank2015Ade}. Recent constraints from the Lyman-alpha forest suggest a very conservative upper limit of $B_0 < 0.3$~nG on a Mpc scale for $n_S = -2.9$ \citep{77rd-vkpz}. Cosmic microwave background (CMB) polarization places limits on $B_0$ in the [0.06, 1.06] nG range for $n_S$ in the [2, -2.9] range \citep{10.1093/mnras/stac2947}. Therefore, in this paper, we considered magnetic fields whose rms strength on the scale of 1 Mpc does not exceed a few nG. We considered two extreme cases in which the magnetic fields are completely non-helical and completely helical.

\section{Thermal history of the early Universe with PMF$\textrm{s}$} 
\label{thermal}
At the epoch of hydrogen recombination in the early Universe, when the viscosity of the cosmic fluid, consisting of baryonic matter and photons, drops sharply, the energy of the magnetic field is converted into heat through damping of magnetohydrodynamic (MHD) modes. This conversion is caused by dissipation processes, which arise from the finite mean free paths of neutrinos and photons, respectively \citep{Jedamzik:1998, Durrer:2013}.
Before hydrogen recombination, there is a tight coupling between photons and baryons through Thomson scattering, which leads to the equality of their temperatures.
After the recombination epoch, the magnetic fields dissipate their energy and heat the IGM gas that is weakly coupled to the radiation. As a result, the evolution of the temperature of the IGM gas $T_\mathrm{b}$ with the heating from the PMFs is given by \citep{Sethi2005}
\begin{equation}
 \frac{dT_\mathrm{b}}{dt} = -2H T_\mathrm{b}
+ \frac{x_\mathrm{e}}{1+x_\mathrm{e}+f_{He}}
\frac{8\rho_\gamma \sigma_\mathrm{T}}{3m_\mathrm{e} c} (T_\gamma - T_\mathrm{b})
+ \frac{2\dot{Q}_\mathrm{mf}}{3k_\mathrm{B} n_\mathrm{b}} ~,
\label{dTgasdt}
\end{equation}
where $m_e$, $x_\mathrm{e}$, $\rho_\gamma$, $T_\gamma$, $n_b$ are electron mass, fraction of electrons, the energy density of CMB radiation,  the radiation temperature, and the number density of baryons, respectively, $\sigma_\mathrm{T}$ is the Thomson cross section, $f_{He} = (1+4(1-Y_{\rm He})/Y_{\rm He})^{-1} \approx 0.077$ for helium mass fraction $Y_{\rm He} = 0.2446$, and $\dot{Q}_\mathrm{mf} = \dot{Q}_\mathrm{AD} + \dot{Q}_\mathrm{DT}$ represents the heating due to two dissipative processes caused by magnetic fields in the weak ionized IGM gas, known as ambipolar diffusion (AD) and decaying turbulence (DT), respectively. The time dependence of the PMFs energy density is given by
\begin{equation}
\frac{d\rho_\mathrm{mf}}{dt}
= -4H\rho_\mathrm{mf} - \dot{Q}_\mathrm{mf},
\label{rho_B}
\end{equation}
where the first term on the right-hand side describes the adiabatic change of PMFs energy density, and $\dot{Q}_\mathrm{mf}$ is the rate of PMFs energy loss due to the dissipation in the plasma. Let us successively evaluate each of the processes, DT and AD.

Turbulence begins to dominate after recombination on scales $k < k_\mathrm{D}^\mathrm{rec}$ due to the drop in radiative viscosity~\cite{Subramanian:1998fn, Jedamzik:1998, Banerjee:2004}. The rate of turbulence energy dissipation into heat is \cite{Sethi2005}
$$
\dot{Q}_\mathrm{DT} = \frac{3w_B}{2} H \rho_{\rm mf}a^4
\frac{\left[\ln(1+t_\mathrm{d}/t_\mathrm{rec})\right]^{w_B}}{\left[\ln(1+t_\mathrm{d}/t_\mathrm{rec}) + \ln(t/t_\mathrm{rec})\right]^{1+w_B}}~,
$$
where $H$ is the Hubble parameter, $a$ is the cosmic scale factor, $t_d = \left(k_{\rm D}V_A\right)^{-1}$ is the physical decay time scale for the turbulence, which can be expressed in relation to the time of recombination $t_{rec}$ as follows $t_d/t_{rec} = 14.8\left(B_0/1\, \textrm{nG}\right)^{-1}\left(k_{\rm D}^{rec}/ 1\, \textrm{Mpc}^{-1}\right)^{-1}$. For the non-helical magnetic fields vorticity $w_B\equiv 2(n_S+3)/(n_S+5)$, while for maximally helical magnetic fields $w_B\equiv 2/3$. 

After recombination, the PMFs caused a relative motion between ions and neutrals in weakly ionized baryonic matter. In a weakly ionized plasma, the timescale of momentum exchange between ions and neutrals is much shorter than the dynamical timescale. As a result, the ion–neutral drift velocity, $\mathbf{u}_i-\mathbf{u}_n$, is set by the balance between the collisional drag and the Lorentz force: $\sum_{in}\xi_{in} \rho_i\rho_n \left(\mathbf{u}_i-\mathbf{u}_n\right) = \mathbf{J}\times \mathbf{B}$. Here, the current density is $\mathbf{J}=(4\pi)^{-1}(\nabla\times \mathbf{B})$, $\rho_i$ and $\rho_n$ denote the ion and neutral mass densities, $\xi_{in} = \langle\sigma v\rangle_{in}/(m_n+m_i)$ is the rate of momentum transfer between ions and neutrals, $\langle\sigma v\rangle_{in}$ is the collision rate coefficient for ion-neutral collisions, and $m_i$ and $m_n$ are the ion and neutral masses, respectively.

Momentum transfer in the early Universe plasma occurs mainly due to collisions between protons and neutrals. The collisions between electrons and neutrals can be neglected, since their contribution is suppressed by the coefficient $m_{\rm e}/m_{\rm n}$, where $m_e$ is the mass of the electron. Hence, protons are the only charged particles to consider and the energy of the magnetic fields is transformed into the thermal energy of the plasma through Joule dissipation at a rate $   \sum_{in}\rho_i\rho_n\xi_{in}\left(\mathbf{u}_i-\mathbf{u}_n\right)^2 \approx \sum_{n}\rho_p\rho_n\xi_{pn}\left(\mathbf{u}_p-\mathbf{u}_n\right)^2$. Since the difference between the velocities of protons and various neutrals is the same, it can be taken under the sum operator, which leads to the expression for the heating rate due to the ambipolar diffusion \citep{Padoan_2000, PhysRevD.70.123003}
\begin{eqnarray}
    \dot{Q}_\mathrm{AD} &\approx& \sum_{n}\rho_p\rho_n\xi_{in}\times\left[\frac{\mathbf{J}\times \mathbf{B}}{\sum_{n}\rho_p\rho_n\xi_{pn}}\right]^2\nonumber\\
    &=& \frac{|(\nabla \times \mathbf{B}) \times \mathbf{B}|^2}{16\pi^2 } \frac{1}{\sum_{n}\rho_p\rho_n\xi_{pn}}.
\end{eqnarray}
Thus, for the plasma of the early Universe, it is sufficient to take into account the coupling between protons and neutral hydrogen and helium atoms, which gives \citep{draine1980interstellar, Shang_2002}
\begin{equation*}
 \sum_{n}\xi_{pn}\rho_n= \frac12n_{\rm H}\langle\sigma v\rangle_\mathrm{pH} + \frac45n_{\rm He}\langle\sigma v\rangle_\mathrm{pHe},
\end{equation*}
where $n_{\rm H}$ and $n_{\rm He}$ are the number densities of neutral atoms of H and He, respectively, and, with a good approximation \cite{PhysRevD.78.083005},
\begin{eqnarray}
    \langle\sigma v\rangle_\mathrm{pH} &\approx& 6.49 \times 10^{-10} (T_\mathrm{b}/\mathrm{K})^{0.375}\, \textrm{cm}^3\textrm{s}^{-1},\nonumber\\
    \langle\sigma v\rangle_\mathrm{pHe} &\approx& (1.424 + 7.438\times10^{-6}T_\mathrm{b} \nonumber\\
&& - 6.734\times10^{-9}T^2_\mathrm{b})\times10^{-9} \, \textrm{cm}^3\textrm{s}^{-1}.\nonumber
\end{eqnarray}
Taking into account that the mass density of protons $\rho_p = x_p\rho_b/(1+Y_{\rm He})$, where $x_p$ is the number fraction of protons, and $\rho_b$ is the mass density of baryons in the Universe, finally, we can obtain that the rate of magnetic field energy dissipation into heat during ambipolar diffusion is 
\begin{equation}
\dot{Q}_\mathrm{AD} = \frac{|(\nabla \times \mathbf{B}) \times \mathbf{B}|^2}{16\pi^2 } \frac{1+ Y_{\rm He}}{\gamma_p\rho_\mathrm{b}^2x_\mathrm{p}} ,
\end{equation}
where $\gamma_p \equiv \sum_{n}\xi_{pn}\rho_n/\rho_b$.

\begin{figure}
\centerline{\includegraphics[width=0.5\textwidth]{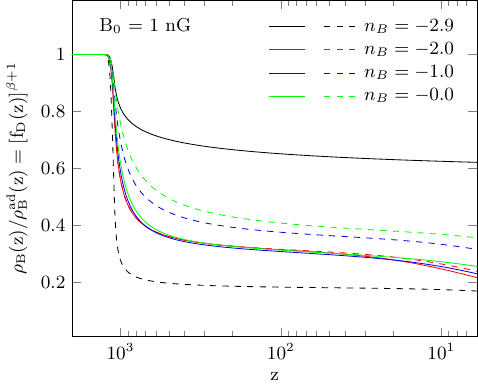} }
\caption{Decreasing PMFs energy density as a result of dissipation processes in the early Universe for different values of power index $n_B$ and for helical (dashed lines) and non-helical (solid lines) PMFs. }
\label{PMFED}
\end{figure}

\begin{figure}
\centerline{\includegraphics[width=0.5\textwidth]{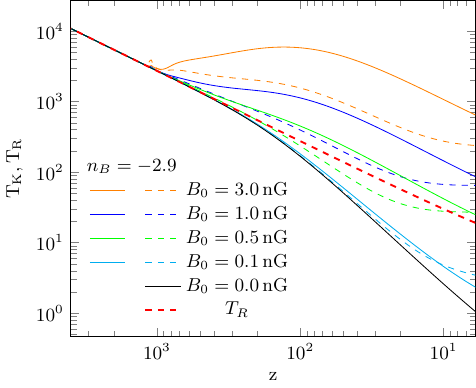} }
\caption{Redshift dependencies of baryonic gas temperature (solid and dashed lines) and CMB temperature (red dashed line)  in cosmological models with PMFs. Solid lines correspond to non-helical PMFs and dashed lines correspond to helical PMFs with rms strengths of $B_0 = 0.1$~nG, $B_0 = 0.5$~nG,  $B_0 = 1.0$~nG, and $B_0 = 3.0$~nG.}
\label{TgTcmb}
\end{figure}

The time dependence of the power spectra of PMFs in the presence of DT and AD can be directly evaluated from 3D MHD simulations \citep{Masson2012, Ntormousi2016}. 
A simpler way is to use self-similarity in the evolution of the magnetic energy spectrum \citep{PhysRevLett.118.055102, PhysRevE.64.056405, PhysRevD.70.123003, Subramanian_2016, PhysRevD.87.083007}
\begin{equation}
    E_M(k,t) \approx f_D^{\beta}(t)\phi(kf_D^{-1}(t)),
\end{equation}
where $E_M(k,t) = 4\pi k^2P_S(k,t)$ is the shell-integrated energy spectrum of PMFs, the function $\phi$ describes the invariant form of the magnetic power spectrum during its evolution, 
$f_\mathrm{D}(t)\equiv k_\mathrm{D}(t)/k_\mathrm{D}^\mathrm{rec}$  is the characteristic change of damping scale of magnetic fields after recombination, $\beta$ is a parameter whose value is some real number that can be set as follows: for maximally helical magnetic fields $\beta\approx 0$, and for non-helical magnetic fields $\beta=n_S+2$. This reflects differences in the redistribution of the kinetic energy of turbulence when a hierarchy of vortices arises. Non-helical fields evolve under a direct cascade of magnetic energy -- energy flows from large to small scales, where it is dissipated. Helical fields undergo an inverse cascade -- magnetic energy shifts from small to larger scales due to the conservation of helicity.

In this approximation, the time dependence of the energy density of PMFs can be expressed through the time dependence of the cosmological scale factor and the damping scale \citep{Minoda2019}: $\rho_\mathrm{mf}(t) = \rho_\mathrm{mf}^\mathrm{rec}\cdot[a(t_{rec})/a(t)]^4\cdot[f_\mathrm{D}(t)]^{\beta+1}$, where $\rho_\mathrm{mf}^\mathrm{rec}\equiv \rho_\mathrm{mf}(t_\mathrm{rec})$ and $f_\mathrm{D}(t)$ can be obtained by integrating the differential equation (\ref{rho_B}) with the initial condition $f_\mathrm{D}(t_\mathrm{rec})=1$. The mean square Lorentz force in this approximation can be represented as follows \citep{Kunze:2013uja}
\begin{eqnarray}
 \langle \mathrm{L}^2 \rangle &\equiv& |(\nabla \times \mathbf{B}) \times \mathbf{B}|^2 \approx 16\pi^2{k^2_\mathrm{D}(t)\rho^2_\mathrm{B}(t)}f_\mathrm{L}(n_B)\hspace{0cm}\nonumber\\
  &=& 16\pi^2\left({k_\mathrm{D}^\mathrm{rec}\rho_\mathrm{mf}^\mathrm{ad}(t)}\right)^2f_\mathrm{L}(n_B)\cdot [f_\mathrm{D}(t)]^{2\beta+4}\cdot a^{-2}(t),\nonumber
\end{eqnarray}
where for non-helical magnetic fields $n_B=n_S$ and $f_\mathrm{L}(n_B)$ can be approximated as
$f_\mathrm{L}^\mathrm{NH}(n_B) \simeq 0.8313\cdot(1 - 1.020\cdot 10^{-2} (n_B + 3)) (n_B + 3)^{1.105}$, while for maximally helical magnetic fields $n_B=n_S=n_H$ and $f_\mathrm{L}^\mathrm{H}(n_B) \simeq 0.45\cdot(1 - 0.017(n_B + 3)) (n_B + 3)^{1.1}$ \citep{Jagannathan:2021}. We suppose that there is no dissipation of PMFs after the epoch of reionization. Therefore, the relationship between the PMFs energy density at the present moment of time $t_U$, $\rho_\mathrm{B}(t_U)$, and the PMFs energy density at the time of recombination $t_{\rm rec}$, $ \rho_\mathrm{mf}^\mathrm{rec}$, is as follows
$$\rho_\mathrm{mf}(t_U) = \rho_\mathrm{mf}^\mathrm{rec}\cdot a^4(t_\mathrm{rec})\left[f_\mathrm{D}(t_\mathrm{rei})\right]^{\beta+1},$$
where $t_\mathrm{rei}$ is the time of the cosmological reionization.

Fig.~\ref{PMFED} shows the decreasing energy density of PMFs as a result of dissipation processes in the early Universe obtained from Eq.~(\ref{rho_B}) for different values of the spectral index $n_B$. Dashed lines correspond to helical PMFs, while solid lines represent non-helical PMFs. The energy density of PMFs is shown as a function of the redshift relative to its adiabatic value $\rho^{\rm ad}_{\rm mf}(z) \equiv \rho^{\rm rec}_{\rm mf}\cdot [(1+z)/(1+z_{rec})]^4$. As can be seen from Fig.~\ref{PMFED}, the relative loss of magnetic field energy due to heating of gas in the early Universe reaches tens of percent and depends on the spectral index and helicity of the PMFs. For non-helical magnetic fields, the loss grows as the spectral index increases, whereas for helical magnetic fields, the loss diminishes with increasing spectral index.

Fig.~\ref{TgTcmb} shows the gas temperature in the early Universe as a function of redshift for cosmological models with helical (colored dashed lines) and non-helical (colored solid lines) PMFs for different values of rms strength at $\lambda = 1$~Mpc. The solid black line shows the temperature of the baryonic gas in the absence of PMFs. The red dashed line shows the radiation temperature. As can be seen from this figure, PMFs with $B_0>0.1$~nG can noticeably change the temperature of the gas in the early Universe, which will affect the rate of chemical reactions and activations of the molecular energy levels. Non-helical PMFs with spectral index $n_B = -2.9$ heat the gas more strongly than helical PMFs of the same strength, although the energy losses of helical PMFs are greater than those of the non-helical ones (see black lines on Fig.~\ref{PMFED}). The reason is that the flow of energy from helical PMFs to baryonic matter occurs mainly at the moment of recombination, when the heating is not so noticeable compared to the thermal energy of the gas. 

\section{Primordial chemistry} \label{pchemistry}

\begin{figure*}
\centerline{\includegraphics[width=1.0\textwidth]{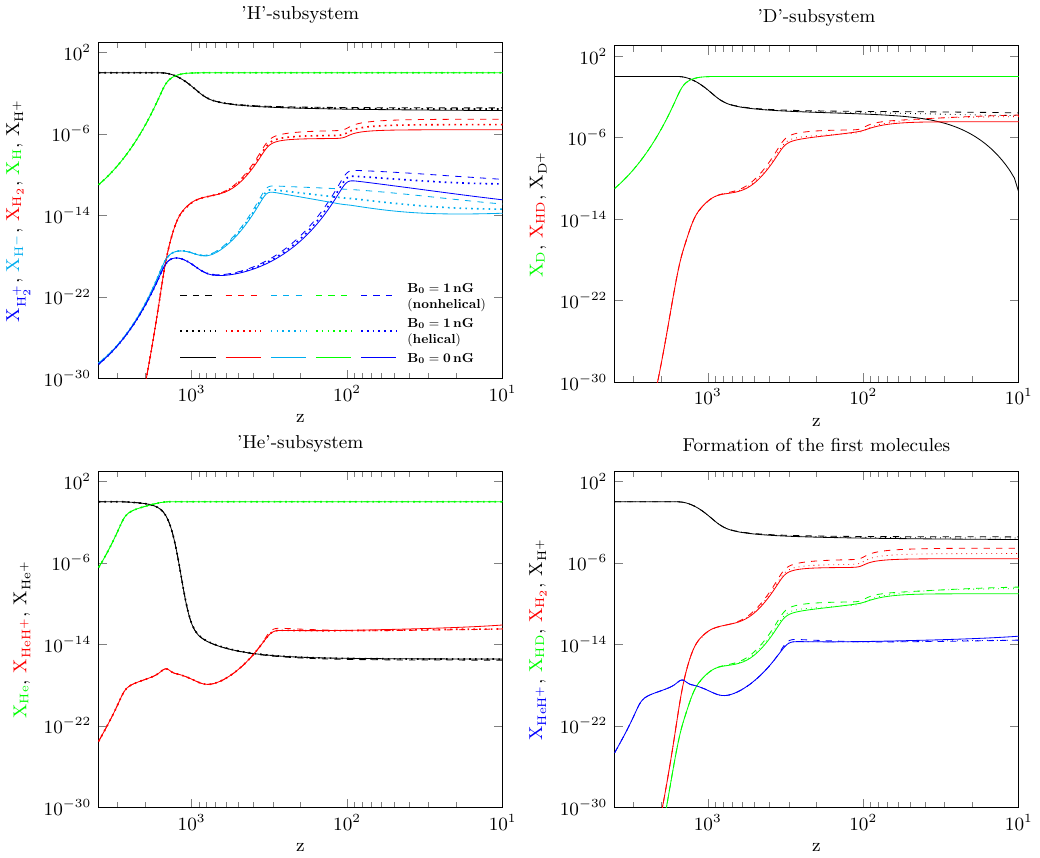} }
\caption{Relative abundances of atoms, molecules, and their ions at different redshifts in cosmological models with (dashed lines for non-helical and dotted lines for helical PMFs with rms strength $B_0 = 1$~nG and spectral index $n_B=-2.9$) and without PMFs (solid lines).}
\label{chemistry}
\end{figure*}
The network of primordial chemistry today may include about 300 reactions for about 30 reagents \citep{bk978-0-7503-1425-1ch1bib38,kulnov2024aa}. However, not all reagents are sufficiently abundant in the early Universe, and not all reactions are important in the conditions of the early Universe. In this paper, we have used a chemical reaction network consisting of 20 reactants and 167 reactions, 166 of which are described in our previous paper \citep{kulnov2024aa}. We supplemented a reaction $\rm HeH^{+} + D \to HD^{+} + He$ and updated the rate for reaction $\rm HeH^{+} + H \to H_2^{+} + He$  according to the latest results of \citep{Grussie2025}. Our chemical network includes reactions involved in the formation of the most common molecules H$_2$, HD, and HeH$^+$. We do not consider reactions involving lithium nuclei because of their small number in the early universe. The network of primordial chemistry can be conditionally divided into subsystems of reactions that are involved in the formation of three two-component molecules: hydrogen molecules H$_2$ \citep{bk978-0-7503-1425-1ch1bib54, bk978-0-7503-1425-1ch1bib58, bk978-0-7503-1425-1ch1bib23, bk978-0-7503-1425-1ch1bib61}, deuterium hydride molecules HD \citep{bk978-0-7503-1425-1ch1bib58, bk978-0-7503-1425-1ch1bib39, bk978-0-7503-1425-1ch1bib38}, and helium hydride ion molecule HeH$^+$ \citep{bk978-0-7503-1425-1ch1bib83, bk978-0-7503-1425-1ch1bib7}. These subsystems are not independent because they include common reagents such as photons, electrons, protons, and hydrogen atoms. The number density of the $i$-th reagent, $n_{\rm i}$, changes as a result of both the expansion of the Universe and chemical reactions. Therefore, the amounts of reagents are usually represented in relation to the total number of basic nuclei, H, D, or He, whose number densities change as $n_\mathrm{H}^\mathrm{tot} \sim n_\mathrm{D}^\mathrm{tot} \sim n_\mathrm{He}^\mathrm{tot}  \sim (1 +z)^3$. The kinetic equation for the fraction of the $i$-th reagent, $X_{\rm i} = n_{\rm i}/n_\mathrm{H}^\mathrm{tot}$, $n_{\rm i}/n_\mathrm{D}^\mathrm{tot}$, or $n_{\rm i}/n_\mathrm{He}^\mathrm{tot}$, can be presented as follows \citep{Puy1993,Galli1998,Novosyadlyj2018}:
\begin{eqnarray*}
\left(\frac{dX_i}{dt}\right)_{\rm{chem}}=\sum_{mn}k^{(i)}_{mn}f_{{m}}f_{{n}}X_{m}X_{n}+\sum_{m}k^{(i)}_{m\gamma}f_{{m}}X_{m}\nonumber\\
-\sum_{j}k_{ij}f_{{i}}f_{{j}}X_{i}X_{j}-k_{i\gamma}f_{{i}}X_{i},
\label{chem_kinetic}
\end{eqnarray*}
where $k_{mn}$ -- reaction rates for the reactants $m$ and $n$,  $f_{{m}}$ or $f_{{n}}$ is $f_{\rm He}=n_{\rm He}/n_{\rm H}$ for the reactants containing He, $f_{\rm D}=n_{\rm D}/n_{\rm H}$ for the reactants containing D, and $f_{\rm H}=n_{\rm H}/n_{\rm H}\equiv1$ for the reactants containing H.

We have integrated the system of kinetic equations for the considered network of chemical reactions in the early Universe. Some of the results are shown in Fig.~\ref{chemistry}, where we compare the kinetics of the formation of the first molecules and their main ingredients in three cosmological models: with helical PMFs  (dotted lines), non-helical PMFs (dashed lines), and in the standard model without PMFs (solid lines). The different panels show the relative abundances of the representative species involved in the formation of H$_2$ (upper left), HD (upper right), and HeH$^+$ (lower left) in units of hydrogen nucleus number density (for the H subsystem), deuterium nucleus number density (for the D subsystem), and helium nucleus number density (He subsystem), accordingly. The lower right panel shows the abundances of the first molecules in units of hydrogen nucleus number density at different redshifts. As expected, additional heating of the baryonic gas by the PMFs leads to an increase in the number of collisions between the reactants. This speeds up the formation and destruction of the first molecules, leading to an increase in the number of H$_2$ and HD molecules and a decrease in the number of HeH$^+$ molecules compared to the case without PMFs. The largest increase in the number of molecules is observed for the hydrogen molecule -- about an order of magnitude for non-helical PMFs and about three times for helical PMFs with rms strength $B_0 = 1$~nG and spectral index $n_B=-2.9$. It is clear that for weaker magnetic fields, the deviations from standard cosmology will be smaller.

\section{Signals from the first molecules of the Dark Ages} 
\label{fmolecules}

\subsection{Rovibrational levels population}

\begin{figure}
 \centerline{\includegraphics[width=0.5\textwidth]{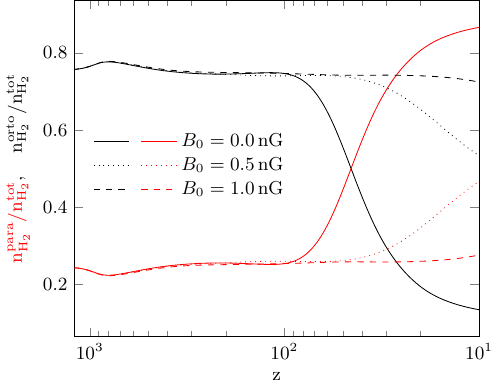} }
\caption{Population of ortho- and para-hydrogen molecules in the early Universe in cosmological models with non-helical PMFs of rms strengths $B_0 = 1.0$~nG (dashed lines) and $B_0 = 0.5$~nG (dotted lines) with spectral index $n_B=-2.9$ compared to a cosmological model without PMFs (solid lines).}
 \label{ortopara}
\end{figure}

\begin{figure*}
 \centerline{\includegraphics[width=1.0\textwidth]{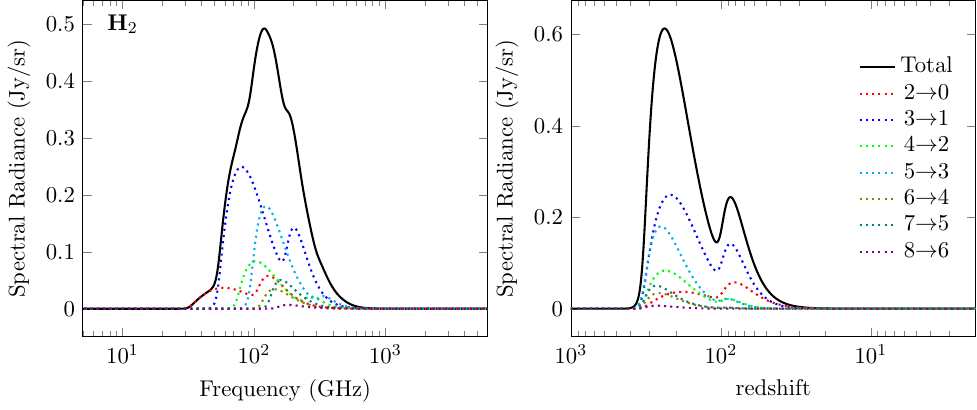} }
 \centerline{\includegraphics[width=1.0\textwidth]{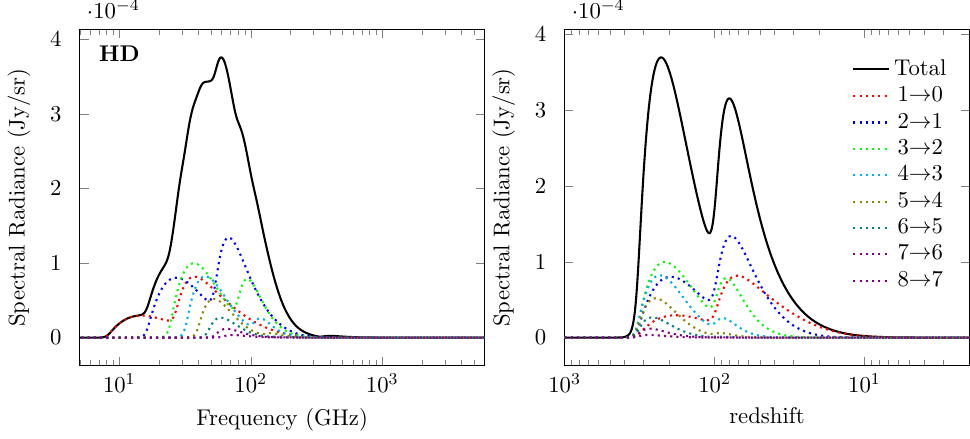} }
 \caption{Emission signals in the rotational lines of molecules -- H$_2$ (upper row) and HD (bottom row) in a cosmological model with nonhelical PMFs with rms magnitude $B_0 = 1$~nG and spectral index $n_B=-2.9$. }
 \label{H2andHDsignals}
\end{figure*}

\begin{figure*}
 \centerline{\includegraphics[width=1.0\textwidth]{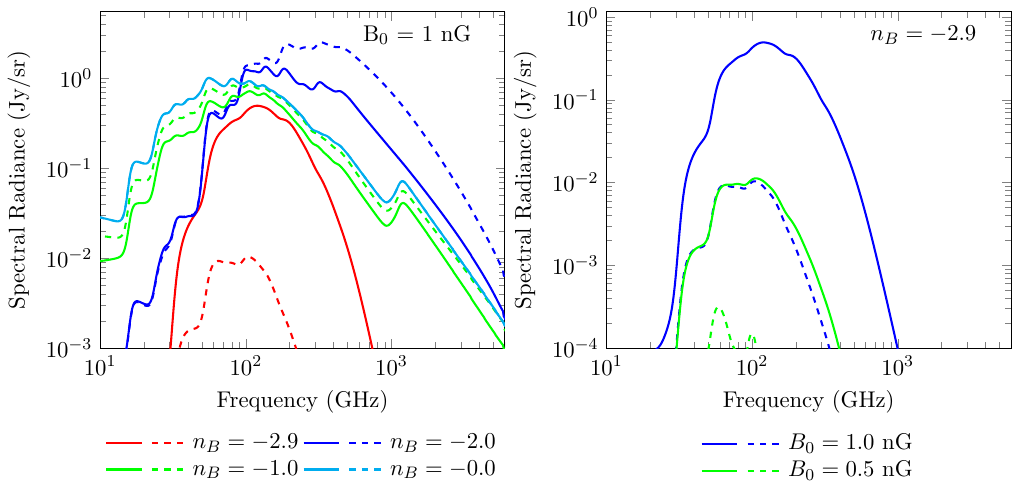} }
 \caption{Combined signal from the first molecules H$_2$ and HD for different values of the spectral index $n_B = -2.9, -2.0, -1.0, 0.0$ (left panel), rms strengths B$_0 = 1.0, 0.5$~nG (right panel), and helicity (dashed lines for fully helical and solid lines for fully nonhelical PMFs). }
 \label{sig_mols}
\end{figure*}

The populations of the rovibrational  levels of H$_2$ and HD molecules in the early Universe are not equilibrial \citep{10.3847/1538-4357/ab530f} and must be described by kinetic equations \citep{kulnov2024aa}
\begin{eqnarray}
 \frac{dX_{v,j}^m}{dt} &=& \sum\limits_{s} n_{s}\left[\sum\limits_{v',j'}k^{s}_{v',j'\to v,j}X_{v',j'}^m-X_{v,j}^m\sum\limits_{v',j'}k^{s}_{v,j\to v',j'}\right] \nonumber\\
 && + \left[\alpha_{v,j}^m\frac{d_{+}}{dt}-X_{v,j}^m\frac{d_{-}}{dt}\right]\ln(n_{m}a^3).
 \label{levels_occupation}
\end{eqnarray}
where $X_{v,j}=n^m_{v,j}/\sum_{v',j'}n^m_{v',j'}$ is the fraction of molecules $m\in\{$H$_2$, HD, HeH$^+\}$ that are in the excited state $j$, and the index s $\in\{\gamma$, H, He, p$\}$ indicates the collision partner of the molecule $m$ that caused the transitions. Here we assume that the transitions between the rovibrational levels of Dark Ages and Cosmic Down molecules are mainly due to the emission/absorption of CMB photons, collisions with neutral hydrogen and helium atoms, and protons, which are responsible for ortho-para transitions in H$_2$. The last term of Eq.~\ref{levels_occupation} describes the fact that the population of rovibrational levels of newly formed molecules, $\alpha_{v,j}^m$, differs from the population of levels of molecules that have already been exposed to the environment, $X_{v,j}^m$. The first is obviously determined by the gas temperature $T_K$, while the CMB temperature, $T_R$, preferably determines the latter. The derivative $d_{+}\ln(n_ma^3)/dt$ determines the rate of formation and $d_{-}\ln(n_ma^3)/dt$ determines the rate of destruction of molecules $m$.

The values of spontaneous rovibrational transitions allowed by the quantum selection rules, transition frequencies, and energy for the low rovibrational energy levels are listed in Table~1 of \citep{kulnov2024aa}, in which the data for para- and ortho- hydrogen molecules H$_2$ are taken from \cite{Roueff:2019}, the data for hydrogen deuteride molecules HD and helium hydride ion HeH$^+$ are taken from \cite{Amaral:2019}. 

The populations of the rovibrational levels of the first molecules in the early Universe were primarily governed by collisions with blackbody cosmic microwave background radiation quanta with a temperature of $T_R$. Intercomponent collisions in a gas whose temperature $T_K$ differed from the radiation temperature $T_R$ led to perturbations of the populations of the rovibrational levels of the first molecules. The most common components of the gas of the early Universe were neutral atoms of hydrogen and helium. 

The deexcitation rates by collisions with neutral hydrogen atoms for H$_2$  molecules are taken from \cite{Lique:2015}, for HD molecules are taken from \cite{Desrousseaux:2022}. We consider de-excitation of H$_2$ and HD by collisions with He using rates from \cite{Nolte2011}, and \cite{Flower1998, Jwiak2024} respectively. 

Collisions of H$_2$ molecules with protons are much rarer in the early Universe, but they contribute significantly to ortho-para transitions. For collisional deexcitation of H$_2$ molecules by protons, we used the rates from the paper \cite{Gerlich:1990}. Excitation transition rates can be related to deexcitation rates as follows:
\begin{equation}
 k_{l\to u}(T_K) = \frac{(2j_u+1)(2I_u+1)}{(2j_l+1)(2I_l+1)}\exp\left\{-\frac{E_u-E_l}{k_{\rm B}T}\right\}k_{u\to l}(T_K),
\end{equation}
where $u$ and $l$ denote the upper and lower rovibrational  levels, respectively, $E_u$ and $E_l$ are their energies, $T_K$ is the temperature of the baryon gas, $I_u$ ($I_l$) is equal to 0 for even $j_u$ ($j_l$), and equal to 1 if $j_u$ ($j_l$) is odd.

In Fig.~\ref{ortopara} we show how the populations of ortho- and para-hydrogen molecules in the early Universe in cosmological models with non-helical PMFs of rms strengths $B_0 = 1.0$~nG (dashed lines) and $B_0 = 0.5$~nG (dotted lines) with spectral index $n_B=-2.9$ compared to the cosmological model without PMFs (solid lines). 
These transitions are caused by collisional interactions between hydrogen molecules and other gas constituents, mainly neutral hydrogen and helium atoms, as well as protons.
The formation of ortho- and para-hydrogen molecules occurs in a ratio of 3:1 according to the ratio of statistical weights of these isomers. Because of collisions and adiabatic cooling of the gas, a moment comes when the concentrations of ortho and para hydrogen equalize, followed by the dominance of the isomer with lower energy. Because the populations of the rovibrational levels of hydrogen molecules are not in equilibrium, the concentration equalization moment cannot be correlated with any fixed gas temperature. However, as shown in Fig.~\ref{ortopara}, gas that is more strongly heated by the PMFs requires a longer time to cool to this moment.

In this work, we calculated the populations of rovibrational levels for molecules H$_2$ and HD. We did not perform calculations for the HeH molecule$^+$ because the concentration and, accordingly, the expected signal from these molecules are much smaller; also, the transition rates induced by collisions with electrons are known only for a few of the lowest rotational levels, which is not enough \citep{kulnov2024aa}.

\subsection{Signals from the first molecules}

The differential brightness of the first molecules on the background of CMB radiation can be expressed as \citep{{kulnov2024aa}}
\begin{equation}
 \delta I_{ul} = \frac{2h\nu^3_{ul}}{c^2}\left[\frac{1}{e^{\frac{h\nu_{ul}}{k_{B}T_{ex}}}-1} - \frac{1}{e^{\frac{h\nu_{ul}}{k_{B}T_{r}}}-1}\right]\tau_{ul},
\end{equation}
where the excitation temperature
\begin{equation}
T^{ul}_{ex} \equiv \frac{E_u-E_l}{k_B}\ln^{-1}\left(\frac{g_uX_l}{g_lX_u}\right),
\end{equation}
and the optical depth in the transition lines $u\to l$ in the approximation of a narrow line is as follows \citep{kulnov2024aa}:
\begin{equation}
\tau_{ul} = \frac{1}{8\pi} \frac{\lambda_{ul}^3A_{ul}}{H(z)}n_{{\rm mol}}\left(X_{l}\frac{g_{u}}{g_{l}}-X_{u}\right),
\end{equation}
where $\lambda_{ul}=c/\nu_{ul}$ is the wavelength in the rest frame, $H(z)$ is expansion rate that depends on the redshift, $n_{\rm mol}$ is number density of molecules. We incorporated these expressions into our CosmoChem code, which integrates the system of differential equations listed above, and computed the differential brightnesses of H$_2$ and HD molecules in their rovibrational transition lines at different redshifts, taking into account the cosmological shift of wavelengths. The results are presented in Fig.~\ref{H2andHDsignals}.
In practice, we will not be able to distinguish the signals from individual lines, because, as shown in Fig.~\ref{H2andHDsignals}, signals originating from different redshifts and transitions overlap, forming a superposition
\begin{equation}
\delta I_{\rm tot} = \sum\limits_{\rm transitions}\delta I_{ul}.
\end{equation}
The brightness in the lines depends not only on the temperature of the gas but also on the number of quanta of CMB radiation corresponding to these lines. For lines with higher energy, the number of relic quanta in the early Universe is greater, as a result of which, as can be seen in Fig.~\ref{H2andHDsignals}, the differential brightness of the lines corresponding to higher transitions decreases rapidly.

The left panels of Fig.~\ref{H2andHDsignals} show emission signals in the rotational lines of molecules -- H$_2$ (upper row) and HD (bottom row) in a cosmological model with PMFs with rms magnitude $B_0 = 1$~nG and spectral index $n_B=-2.9$. Frequencies are presented in the reference frame of an observer on Earth. The colored dot lines show contributions from separate transitions, and the black line shows their superposition. The right panels of  Fig.~\ref{H2andHDsignals} show the redshifts at which emission occurred for the corresponding transitions, for H$_2$ (upper row) and for HD (bottom row). For both molecules, two peaks are observed on the redshift scale. They correspond to the positions of the step growth in the number densities of the H$_2$ and HD molecules in Fig.~\ref{chemistry}.  The molecular signal amplitude increases with the growing temperature difference between the gas and the radiation and with the rising molecular abundance. At lower redshifts, the signal diminishes as the expansion of the Universe reduces the collision rate in the gas.

Figure~\ref{sig_mols} shows how the combined signal of the first molecules, H$_2$ and HD, depends on the spectral index, the rms strength, and the helicity of the PMFs. The left panel illustrates the dependence on the spectral index and helicity, while the right panel shows the dependence on the rms strength B$_0$ and helicity. In both panels, dashed lines correspond to fully helical PMFs, and solid lines to fully non-helical PMFs. The graphs show significant sensitivity of the signal amplitude and its spectral range to these parameters.

\subsection{Comparison with other CMB spectral distortions}

\begin{figure}
 \centerline{\includegraphics[width=0.5\textwidth]{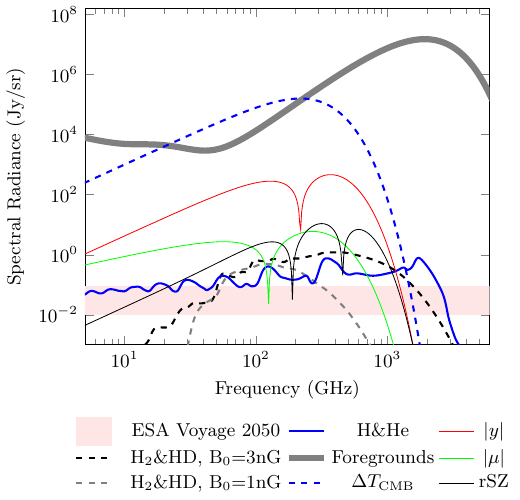} }
 \caption{
The signals from the first molecules H$_2$ and HD (black for $B_0=3$~nG and gray for $B_0=1$~nG dashed lines) compared to expected CMB spectral distortions -- $\mu$--distortion (green solid line),  $y$--distortion (red solid line),  relativistic correction to  $y$--distortion (black solid line), CMB temperature uncertainty (blue dashed line), and the cosmological recombination radiation \citep{10.1093/mnras/stv2691} (blue solid line). Total astrophysical foregrounds are shown with a gray solid line. The pink-shaded area indicates the expected sensitivity of the ESA Voyage 2050 mission.
 }
 \label{distorts}
\end{figure}

The CMB we observe today is extremely close to a blackbody with a temperature of 2.725 K, but various physical processes in the Universe could inject or absorb energy, and thereby distort the spectrum. In particular, the signals from the first molecules H$_2$ and HD fall within the CMB frequency range and contribute to its spectral distortions, manifesting as deviations from a perfect blackbody spectrum. In Fig.~\ref{distorts} we compare signals from the first molecules in the presence of PMFs with others known types of CMB spectral distortions \citep{10.1093/mnras/stw945}: cosmological recombination radiation (blue solid line), total foregrounds (gray dashed line), uncertainty in the temperature of CMB (blue dashed line), y-distortion (red solid line), $\mu$-distortion (green solid line), and relativistic Sunyaev-Zeldovich effect (black solid line). Detecting these distortions will allow us to discover and study several physical processes in the early Universe. Therefore, there are several proposals for mission designs (e.g., PIXIE, super-PIXIE, PRISTINE, PRISM, SIMBAD, LiteBIRD extension, and Voyage2050 \citep{Chluba2021, Maillard2020}) aimed at detecting spectral distortions of the CMB with a higher level of precision than COBE/FIRAS did this more than three decades ago.

In the standard cosmology model, the global signal from the first molecules appears as an absorption feature in the CMB spectrum, with a small amplitude of about $10^{-3}$~Jy/str \citep{kulnov2024aa}, well below the sensitivity of currently planned observational missions. However, the presence of PMFs with a strength of $\sim 1$~nG transforms this signal into emission with an amplitude of $\sim 0.5$~Jy/str (see Fig.~\ref{H2andHDsignals}). As shown in Fig.~\ref{distorts}, this level is comparable to the cosmological recombination radiation (blue solid line) and lies within reach of some proposed missions, such as super-PIXIE  \citep{kogut2025primordialinflationexplorerpixie}, multi-SIMBAD (4-units) \citep{Maillard2020}, and Voyage2050. The black and gray dotted lines in Fig.~\ref{distorts} represent the signals from the first molecules, H$_2$ and HD, in the presence of non-helical PMFs with $n_B = -2.9$ and rms strengths of $B_0 = 3$~nG and $B_0 = 1$~nG, respectively. As we have shown in Fig~\ref{sig_mols}, the emission signal from the first molecules is sensitive to the PMFs parameters. Therefore, future detection of CMB distortions will allow us to determine or constrain their values. Moreover, detecting the signal of the first molecules from the Dark Ages would represent a major milestone in cosmology, providing an additional piece in the overall picture of the Universe.

\section{Conclusions}
\label{concl}
In this work, we have studied the effect of the PMFs on the Universe's thermal history between the hydrogen recombination and the reionization epoch. We investigated how the presence of PMFs affects the chemical dynamics of the first molecules and the occupation of their rovibrational levels. The dissipation of the PMFs energy heats the baryonic gas, which increases the temperature of the baryonic matter relative to the case without PMFs. This, in turn, alters the concentration of first molecules as well as the population of their rovibrational levels. We have carefully considered the processes that transfer the magnetic fields' energy to the plasma (ambipolar diffusion and decaying turbulence), and hence how they change the kinetics of gas temperature. 

Then, with numerical solutions of differential equations that govern the thermal and chemical evolution of the Universe in the considered period, we have found the impact of these processes on the gas temperature during the Hubble expansion for cosmological models with helical and non-helical PMFs for different values of spectral index $n_B$ and rms strength $B_0$ averaged at scale $\lambda = 1$~Mpc. We have tracked how the abundances of H$_2$, HD, and HeH$^+$ molecules, H, D, and He atoms, and their ions between redshifts $\sim$1200 and 10 in the presence of PMFs. We demonstrate that PMFs can significantly accelerate the formation and destruction of the first molecules, resulting in an increase in the number of H$_2$ and HD molecules and a decrease in the number of HeH$^+$ molecules compared to the case without PMFs. The most significant changes occur for the hydrogen molecule. Its concentration increases by nearly an order of magnitude for non-helical PMFs and by about a factor of three for helical PMFs with an rms strength of $B_0 = 1$ nG, and spectral index $n_B = -2.9$, compared with the standard cosmological model. 

We numerically integrate the differential equations governing the kinetics of the rovibrational levels populations of the first molecules (H$_2$ and HD) during the considered epoch of the Universe. We have tracked the population of rovibrational levels of H$_2$ and HD molecules in the presence of PMFs. We showed how the abundance of ortho and para hydrogen changes in a universe with PMFs compared to a universe without PMFs. We conclude that collisions in the gas alter the ortho-to-para ratio of hydrogen molecules, making it a potential probe of the thermal history of the gas in the early Universe.

The signals from the first molecules, H$_2$ and HD, were computed as their differential brightness against the cosmic microwave background. These signals lie within the CMB frequency range and appear as its spectral distortions, manifesting as small deviations from a perfect blackbody spectrum. 

In the standard cosmology, the global signal from the first molecules appears as an absorption feature in the CMB spectrum, with a small amplitude of about $10^{-3}$~Jy/str \citep{kulnov2024aa}. At the same time, the presence of non-helical PMFs with $n_B=-2.9$ and rms strength of $\sim 1$~nG transforms this signal into emission with an amplitude of $\sim 0.5$~Jy/str. We conclude that this level is comparable to the cosmological recombination radiation and is within the reach of some proposed missions, such as super-PIXIE \citep{kogut2025primordialinflationexplorerpixie}, multi-SIMBAD (4-units) \citep{Maillard2020}, and Voyage2050 \footnote{https://www.cosmos.esa.int/web/voyage-2050}. We show that the amplitude and spectral range of the signal are highly sensitive to the spectral index $n_B$, rms strength $B_0$, and helicity of PMFs. Thus, the global signals of the first molecules represent a potential probe of primordial magnetic fields.

\begin{acknowledgements}
This work is done in the framework of the project \emph{“Tomography of the Dark Ages and Cosmic Dawn in the lines of hydrogen and the first molecules as a test of cosmological models”} (state registration number 0124U004029) supported by the National Research Foundation of Ukraine.
\end{acknowledgements}
\newpage
\bibliographystyle{h-physrev}
\bibliography{arefs}

\end{document}